\documentclass[floatfix,aps,amsmath,amssymb,reprint]{revtex4-2}

\usepackage{graphicx}
\usepackage{dcolumn}
\usepackage{bm}
\usepackage{placeins}
\usepackage{float}
\usepackage{appendix}
\usepackage{xcolor}
\begin{document}

\title{Activity-driven tissue alignment in proliferating spheroids}%

\author{Liam J. Ruske}
\email{liam.ruske@physics.ox.ac.uk}
\author{Julia M. Yeomans}
\affiliation{Rudolf Peierls Centre For Theoretical Physics, University of Oxford, UK}

\date{\today}

\begin{abstract}
We extend the continuum theory of active nematic fluids to study cell flows and tissue dynamics inside multicellular spheroids, spherical, self-assembled aggregates of cells that are widely used as model systems to study tumour dynamics. Cells near the surface of spheroids have better access to nutrients and therefore proliferate more rapidly than those in the resource-depleted core. Using both analytical arguments and three-dimensional simulations, we find that the proliferation gradients result in flows and in gradients of activity both of which can align the orientation axis of cells inside the aggregates. Depending on environmental conditions and the intrinsic tissue properties, we identify three distinct alignment regimes:  spheroids in which all the cells align either radially or tangentially to the surface throughout the aggregate and  spheroids with angular cell orientation close to the surface and radial alignment in the core. 
The  continuum description of tissue dynamics inside spheroids not only allows us to infer dynamic cell parameters from experimentally measured cell alignment profiles, but more generally motivates novel mechanisms for controlling the alignment of cells within aggregates which has been shown to influence the mechanical properties and invasive capabilities of tumors.
\end{abstract}


\maketitle

\section{\label{sec:intro}Introduction}

Many biological systems such as epithelial tissues, bacterial colonies or three-dimensional organoids are active systems; they continuously consume chemical energy from their environment to perform tasks such as growth or migration, both of which involve the generation of mechanical stresses by the cells through a variety of biochemical processes \citep{alberts2008molecular, pollard2003cellular, dogterom2005force}.  
A prominent 3D example of an active tissue is multicellular spheroids. These are spherical, self-assembled aggregates of cells which provide important model systems for investigating the behaviour of 3D cell aggregates, for example, the effects of mechanical stress, or for screening of anti-cancer drugs.

Multicellular spheroids are typically studied in a suitable environment which provides cells with all the needed metabolites such as oxygen and nutrients. The metabolite concentration within cellular aggregates therefore normally decreases towards the core since the supply of metabolites in spheroids and avascular tumors is limited by diffusive transport from the surface. While cells close to the surface show a high rate of cell division, cells located close to the core do not have sufficient access to metabolites and die, thereby forming a necrotic core \citep{hirschhaeuser2010multicellular, jagiella2016inferring}. Hence the outer shell of spheroids is dominated by cell growth while the core is dominated by cell death. This has two consequences. The first is that a radial flow of cells is set up from the surface to the centre of the spheroid. The second is a gradient of activity which is a consequence of the gradient in the number of cell divisions which locally generate active, extensile, dipolar forces along their division axes. In this paper we study the competition between these effects to describe the flow and orientation of cells in dense multicellular aggregates.

Our results are based on the continuum theory of three-dimensional active nematic fluids, which has been shown to reproduce the spatio-temporal dynamics in a variety of living systems, from myosin-driven actin-microtubule networks \citep{lee2021myosin} to bacterial biofilms \citep{yaman2019emergence} and confluent cell monolayers \citep{saw2017topological, mueller2019emergence, duclos2017topological, saw2018biological}. 
We start in section \ref{sec:methods} by presenting the hydrodynamic theory that describes cell flows and cell orientation inside dense cell aggregates and list the equations of motion which we solve analytically and numerically. The main results are presented in section \ref{sec:results}, which is divided into four subsections: 

In the first part, \ref{subsec:flows}, we show that radial cell flows created by gradients in the isotropic cell growth rate give rise to radial cell elongation profiles inside multicellular spheroids, where cells are aligned either radially or circumferentially (tangentially) throughout the aggregate. 

In the second part \ref{subsec:active}, we investigate how anisotropic active forces produced by cell division and death affect the cell orientation and flows inside spheroids. We highlight that gradients in active stress give rise to an effective cell anchoring at the surface and in the bulk, and the emergence of chaotic cell flows in the aggregate if active stress is sufficiently large.

We then show, in \ref{subsec:flowact}, that the interplay between growth driven flow and anisotropic stress can give rise to a variety of cell alignment profiles in proliferating spheroids. We subsequently demonstrate, in \ref{subsec:infer}, that our theory allows us to estimate mechanical and dynamical tissue parameters directly from experimental measurements of the distribution of cell orientations inside freely grown spheroids. 

The last section of the paper, \ref{sec:conclusion}, summarizes the key results and points out future research directions that it would be interesting to pursue as more sophisticated experimental data becomes available.

\section{\label{sec:methods} Theoretical Model}


We consider a multicellular spheroid embedded in an environment with a fixed concentration $m_0$ of metabolites, such as oxygen, glucose or growth factors, which are necessary for cell proliferation. We assume that metabolites freely diffuse throughout tissue with diffusion constant $D_m$ and are consumed by cells at a rate proportional to $\alpha^\star$. In the diffusive limit (Péclet number $Pe \ll 1$), the time evolution of metabolite concentration $m(\mathbf{x},t)$ across tissues is governed by the reaction–diffusion equation 
\begin{equation}
    \partial_t m = D_m \Delta m - \alpha^{\star} \varphi \: m\: ,
    \label{eq:cdiff}
\end{equation}
where $\varphi(\mathbf{x},t)$ is the cell-number density and $\Delta$ is the Laplace operator. After an initial growth phase, multicellular spheroids reach a steady state at a finite radius $R_c$ in which cell division near the outer shell exactly balances cell death close to the core. At steady-state, the cell density inside multicellular spheroids is homogeneous \citep{delarue2014stress, montel2012isotropic, alessandri2013cellular}, so we may write $\varphi(\mathbf{x},t)=\varphi_0$ and define $\alpha=\alpha^{\star} \varphi_0$ as a constant metabolite consumption rate throughout the aggregate. Since metabolites diffuse much faster than cellular rearrangements in tissues or aggregate deformations, the metabolite concentration is always at steady-state and eqn.~(\ref{eq:cdiff}) simplifies to $\Delta m = (\alpha/D_m) m$. Thus $m(\mathbf{x})$ decreases towards the core of spheroids with a characteristic penetration length $\ell_m = \sqrt{D_m/\alpha}$.

We describe the motion of cells inside spheroids in a continuum limit, where tissue is characterised by a cell-velocity field $\mathbf{u}$ and a tensorial order parameter $\mathbf{Q}$, which describes the averaged, local magnitude $S$ and direction $\mathbf{n}$ of cell alignment
\begin{equation}
    Q_{ij} = \frac{3 S}{2} \left(n_i n_j - \delta_{ij}/3\right) \: .
\end{equation}
Since the cell density stays constant throughout steady-state spheroids, the cell flow $\mathbf{u}$ follows the continuity equation of an incompressible fluid
\begin{equation}
    \mathbf{\nabla} \cdot \mathbf{u} = k_p^\star \: ,
    \label{eq:cont}
\end{equation}
where the net cell production rate $k_p^\star(\mathbf{x},t)=k_+ - k_-$ depends on the local reproduction rate $k_+$ or death rate $k_-$ of individual cells. When cells have access to sufficient metabolites, cell division dominates, $k_p^\star > 0$, which acts as a mass source causing diverging local flow. On the other hand, when the metabolite concentration falls below a certain threshold, $m < m_c$, cells die faster than they divide, $k_p^\star<0$, thus creating a mass sink and locally converging flow. In the vicinity of the homeostatic equilibrium, $m=m_c$, we expand the cell production rate $k_p^\star$ rate around the homeostatic metabolite concentration to linear order,
\begin{equation}
	k_p^\star = k_p \left( m - m_c \right) \: ,
	\label{eq:kp}
\end{equation}
where the parameter $k_p$ is a proportionality constant. 

In addition to continuity equation~(\ref{eq:cont}), the cell flow $\mathbf{u}$ in tissues must conserve momentum as described by the Navier-Stokes-equations
\begin{equation}
	\rho \left( \partial_{t} + \mathbf{u} \cdot \mathbf{\nabla} \right) \mathbf{u} = \mathbf{\nabla} \cdot \Pi , \:
	\label{eq:NSE}
\end{equation}
where $\rho$ is mass density. Unlike for 2D cell-monolayers, where friction between cells and the underlying substrate gives rise to a friction term in eqn.~(\ref{eq:NSE}), for 3D spheroids we assume that cell-cell friction is dominating and that friction forces at the boundary of freely grown aggregates are negligible. The stress tensor $\Pi = \Pi_{passive} + \Pi_{active}$ consists of a passive and active contribution, where the passive terms $\Pi_{passive}$ are well known from liquid crystal hydrodynamics \citep{marenduzzo2007hydrodynamics} and presented in appendix~\ref{appx:eom}. In addition, each division and death event contributes anisotropic dipolar forces to the active stress. Assuming that the cell axis is on average aligned with the local tissue anisotropy $\mathbf{Q}$, the active stress due to cell division and cell death is \citep{doostmohammadi2015celebrating, simha2002hydrodynamic}
\begin{equation}
	\Pi_{act} = - \zeta^\star \mathbf{Q} \: ,
	\label{eq:activestress}
\end{equation}
where $\zeta^\star(\mathbf{x},t)$ is the magnitude of active stress along the local elongation axis of cells $\mathbf{n}$). Each cell division produces extensile anisotropic stress, $\zeta_+>0$, where forces are directed outwards along cell axis $\mathbf{n}$ and inwards along the two perpendicular axes. The force direction is reversed for cell death events, which produce contractile stress, $\zeta_-<0$. For simplicity we assume $\zeta_+ = |\zeta_-|$, hence the magnitude of active stress follows variations in cell production rate,
\begin{equation}
	\zeta^\star = \zeta \left( m-m_c \right)  \: ,
	\label{eq:zetarad}
\end{equation}
where $\zeta$ quantifies the local, time-averaged magnitude of active stress induced by cell division or death. Note that, for simplicity, we have neglected the intrinsic active stress produced by cells even in the absence of cell division or death, which can be either extensile or contractile, depending on cell type \citep{blanch2018turbulent,saw2017topological,duclos2017topological}.

Since cells adhere to each other, the shape and orientation of cells in a tissue responds to the cellular flows around them. The time evolution of the nematic tensor $\mathbf{Q}$ is thus coupled to the velocity field $\mathbf{u}$ and follows the Beris-Edwards equation \citep{beris1994thermodynamics}
\begin{equation}
    D_t \mathbf{Q} -\mathbf{\mathcal{W}} = \Gamma \mathbf{H} \: ,
    \label{eq:Qt}
\end{equation}
where $D_t$ denotes the material derivative and $\mathbf{\mathcal{W}}$ is the co-rotational term, which describes how cell orientation responds to gradients in flow field $\mathbf{u}$ \citep{marenduzzo2007steady, marenduzzo2007hydrodynamics}. The last term is the rotational diffusivity $\Gamma$, which controls the relaxation of cell deformations towards the equilibrium cell shape quantified by the molecular field $\mathbf{H} = -\delta \mathcal{F}/\delta \mathbf{Q} + (\mathbf{I}/3) \text{Tr}( \delta \mathcal{F} / \delta \mathbf{Q})$. The free energy $\mathcal{F}$ of tissues encodes the mechanical and geometric properties of cells, such as the aspect ratio at equilibrium and the elastic energy associated with cell deformations (Fig.~\ref{fig1} d). Isotropic cells, such as most epithelial cells, are on average isotropic at equilibrium, hence the minimum of the free energy is at $S_{eq}=0$. Cells like fibroblasts or rod-shaped bacteria, however, exhibit an elongated morphology at equilibrium modelled by choosing $S_{eq}>0$. Expressions for the co-rotational term $\mathbf{\mathcal{W}}$ and the tissue free energy $\mathcal{F}$ are presented in appendix \ref{appx:eom}. 

We simulate growing spheroids by numerically solving the continuum equations of motion eqn.~(\ref{eq:cdiff}) for the metabolite concentration and eqns.~(\ref{eq:cont}), (\ref{eq:NSE}), (\ref{eq:Qt}) for the tissue dynamics, on a three-dimensional grid of size $84 \times 84 \times 84$ using a hybrid lattice Boltzmann-finite difference method \citep{marenduzzo2007steady}. Numerical details and simulation parameters are stated in appendix \ref{appx:eom}. Analytical solutions for the flow field $\mathbf{u}$ and metabolite concentration $m$ in spherical aggregates are presented in the following section \ref{subsec:flows}.

\section{\label{sec:results} Results}

\subsection{\label{subsec:flows} Converging flows promote cell alignment}

 \begin{figure*}[!ht]
	\centering
	\includegraphics[width=16cm]{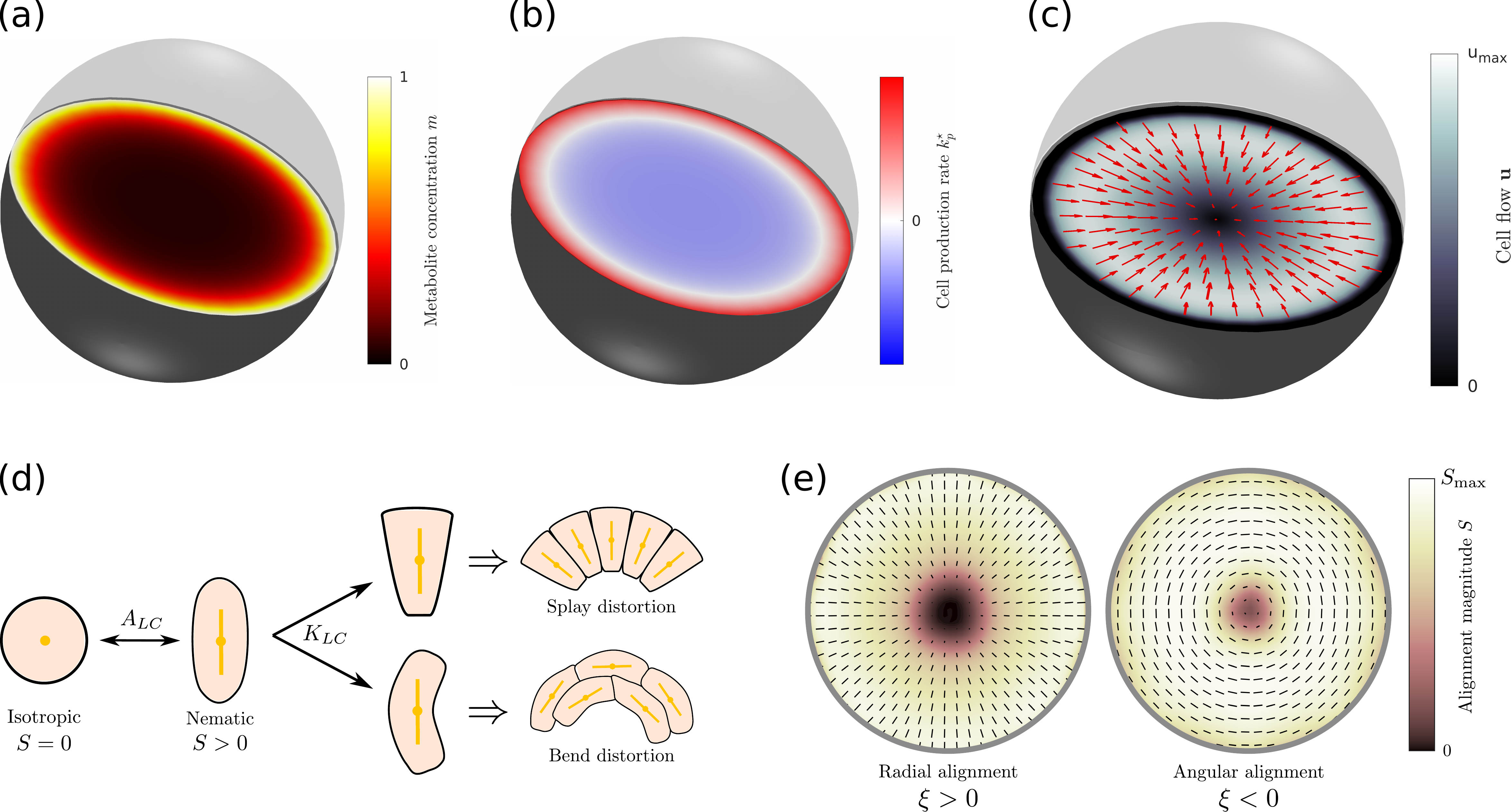}
	\caption{\textbf{(a)} Diffusion limited transport of metabolites give rise to radial concentration profiles $m(r)$ which decrease towards the core of spheroids. \textbf{(b)} Assuming the cell proliferation and cell death rates are proportional to the local metabolite concentration, the cell production rate $k^\star_p$ at steady-state is positive close to the surface and negative in the core, where the transition point $k^\star_p=0$ depends on the critical metabolite concentration $m_c$. \textbf{(c)} In the absence of active dipolar stress spatial variations in the cell production rate create converging flow profiles given by eqn.~(\ref{eq:ur}). \textbf{(d)} The mechanical and geometric properties of cells give rise to a free energy $\mathcal{F}$ which includes elastic constants associated with cell deformations, where $A_{LC}$ penalizes cell stretching which changes the aspect ratio of cells and $K_{LC}$ penalizes changes in cell shape which results in bend and splay deformations. \textbf{(e)} Isotropic cells inside proliferating spheroids are radially elongated by converging flows for alignment parameter $\xi>0$ and align tangential to the surface for $\xi<0$. Since the magnitude of flow-induced elongation increases along the radial direction, cells close to the surface are more elongated than cells in the centre.}
	\label{fig1}
\end{figure*}

We start by analytically solving the equations of motion (\ref{eq:cdiff}), (\ref{eq:cont}), (\ref{eq:Qt}) for the metabolite concentration $m$, cell flow $\mathbf{u}$ and nematic tensor $\mathbf{Q}$ for spherical aggregates in which active dipolar stress produced by cell division and cell death is negligible, $\zeta = 0$. 
In this case, the cell flows are caused solely by mass sources and sinks which arise as a consequence of the cell production rate $k_p^\star$. Solving eqn.~(\ref{eq:cdiff}) for the boundary condition $m=m_0=1$ at the surface yields a metabolite concentration $m(r)$ which is a function of the radial coordinate only,
\begin{equation}
    m(r) = \frac{R}{r} \frac{\text{sinh}(r/\ell_m)}{\text{sinh}(R/\ell_m)} \: ,
    \label{eq:msphere}
\end{equation}
where $R$ is the radius of the spheroid and $\ell_m$ is the characteristic penetration length of metabolites from the surface (Fig.~\ref{fig1} a,b).  If the initial spheroid size is small, $R \lesssim \ell_m$, the total cell production rate is positive and the aggregate grows at a rate 
\begin{equation}
    \dot{R} = k_p \ell_m \left[ \text{coth}(R/\ell_m) - \frac{\ell_m}{R} - \frac{m_c R}{3 \ell_m} \right] \: .
    \label{eq:Rdot}
\end{equation}
As less metabolites reach the core of larger aggregates, spheroids eventually reach a steady state $\dot{R}\rightarrow 0$ at a finite size $R=R_c$. At steady-state, the dimensionless radius $\Tilde{R_c}=R_c/\ell_m$ is a function of the critical metabolite concentration only, $\Tilde{R_c} = f^{-1}(m_c)$, where $f(x)=3(x \coth x - 1)/x^2$ (Fig.~\ref{figS1} a). Fig.~\ref{figS1} b,c shows how $m_c$ affects the shape of the metabolite profile $m(r)$ which determines the cell production rate $k_p^\star$ inside aggregates of size $R=R_c$.

The cell flow $\mathbf{u}=u_r(r) \hat{\mathbf{r}}$ can be obtained by solving the continuity equation (\ref{eq:cont}) in the domain $r\in[0,R]$ using the boundary condition $\mathbf{u}(R)=\dot{R}$ ,
\begin{equation}
     u_r = k_p R \left[ \frac{1}{\Tilde{r}} \frac{\text{cosh}~\Tilde{r}}{\text{sinh}~\Tilde{R}} - \frac{1}{\Tilde{r}^2} \frac{\text{sinh}~\Tilde{r}}{\text{sinh}~\Tilde{R}} - \frac{m_c r}{3 R}  \right]\: ,
     \label{eq:ur}
\end{equation}
where $\Tilde{R} = R/\ell_m$ and $\Tilde{r} = r/\ell_m$. Since $\dot{R}>0$ in growing spheroids, there is a transition between a diverging cell flow close to the surface, $u_r>0$, and a converging cell flow $u_r<0$ in the core of aggregates (Fig.~\ref{figS2} a). At steady-state, cell flows are facing inwards everywhere and vanish at the surface, $u_r(R_c)=0$ (Fig.~\ref{fig1} c), and the shape of flow profiles depends on the critical metabolite concentration $m_c$ (Fig.~\ref{figS1} d). Recently it has been reported that shear stresses created by cell flow $u_r$ can create a viscocapillary instability that perturbs the spherical symmetry of aggregates for sufficiently small surface tension \citep{martin2021viscocapillary}. In this work, however, we consider the limit in which cell-generated stress is small compared to the surface tension of spheroids and the shape of cell aggregates remains spherical.

We now describe how the radial flows $u_r$, given by eqn.~(\ref{eq:ur}), affect the elongation and alignment of cells as quantified by the nematic order parameter $\mathbf{Q}$. To this end, we consider an aggregate of initially isotropic cells, $\mathbf{Q}(\mathbf{x},t)=0$. As outlined in appendix \ref{appx:stbanly}, the stability of the isotropic state is governed by
\begin{equation}
     \dot{Q}_{ij} = \frac{2}{3} \xi \Tilde{E}_{ij}\: ,
     \label{eq:Qst}
\end{equation}
where $\Tilde{E}_{ij} = E_{ij} - \delta_{ij} E_{kk}/3 $ is the traceless part of the strain rate tensor $E_{ij}=(\partial_j u_i + \partial_i u_j)/2$.
The \textit{flow-alignment parameter} $\xi$ describes how pure shear flows affect the orientation of cells, where the sign of $\xi$ determines whether cells align parallel or perpendicular to the shear axis. 

In passive liquid crystals the value of $\xi$ is determined by the aspect ratio of the constituents, which predicts $\xi>0$ for elongated prolate-shaped cells. In the context of tissue dynamics, an alternative flow-alignment parameter $\nu$ is occasionally used \citep{aigouy2010cell, duclos2018spontaneous}, which quantifies the time-evolution of the cell orientation axis $\mathbf{n}$ rather than the nematic tensor $\mathbf{Q}$, where the parameters $\nu$ and $\xi$ are related via $\nu = -\xi (3S+4)/(9S)$. To the best of our knowledge, there are only few studies that have attempted to directly extract the flow-alignment parameter of living tissues. The analysis of cell orientations and flows in confluent layers of spindle-shaped cells under confinement revealed, that tissue dynamics follows $\nu<0$ \citep{duclos2018spontaneous}. Surprisingly, in the wing epithelium of Drosophila cell orientation is differently affected by shear flows in healthy wings ($\nu<0$) and severed wings ($\nu>0$) \citep{aigouy2010cell}. Based on these observations we assume that the flow-alignment parameter $\xi$ is an intrinsic tissue property which may depend on cell type, biochemical signals or the mechanical properties of the environment.

%

Using spherical coordinates, the initial growth rates of the radial component $Q_{rr}$ and angular components $Q_{\phi \phi}$, $Q_{\theta \theta}$ are (see appendix \ref{appx:stbanly})
 \begin{align}
    \label{eq:Qugr}
     \dot{Q}_{rr} &= \frac{4}{9} \xi \left( \partial_r u_r - \frac{u_r}{r} \right)\: , \\
     \dot{Q}_{\theta \theta} &= \dot{Q}_{\phi \phi} = - \frac{1}{2} \dot{Q}_{rr}\: .   
     \label{eq:Qrr}
\end{align} 
From eqn.~(\ref{eq:ur}) it follows that $(\partial_r u_r - u_r /r) \sim \Tilde{r}^{-1} \sinh \Tilde{r} + 3~\Tilde{r}^{-2} ( \Tilde{r}^{-1} \sinh \Tilde{r} -\cosh \Tilde{r}) > 0$, thus isotropic cells inside spheroids become radially elongated for $\xi>0$ and radially compressed for $\xi<0$ (Fig.~\ref{fig1} e). Since the strength of the alignment depends on the magnitude of the strain rate, the growth rate of radial alignment scales as $\dot{Q}_{rr} \sim \xi k_p$ and increases along the radial direction, reaching a maximum at the surface $r=R$. This is shown in Fig.~\ref{figS1} e for different values of $m_c$. Although the magnitude of alignment $S$ initially follows eqn.~(\ref{eq:Qugr}), at later times the director field is subject to higher-order effects such as advection, and flow-induced cell elongation through the co-rotational term $\mathcal{W}$ is eventually balanced by passive restoring forces arising from the molecular field $\mathbf{H}$ (Fig.~\ref{figS1} f). The cell alignment profile at steady state depends on mechanical cell parameters (Fig.~\ref{fig1} d), however, alignment profiles retain the main features seen at early times with cell alignment being maximal near the surface and decreasing towards the core (Fig.~\ref{figS1} g,h).

\subsection{\label{subsec:active} Active stress gradients induce nematic alignment}

 \begin{figure*}[!ht]
	\centering
	\includegraphics[width=16cm]{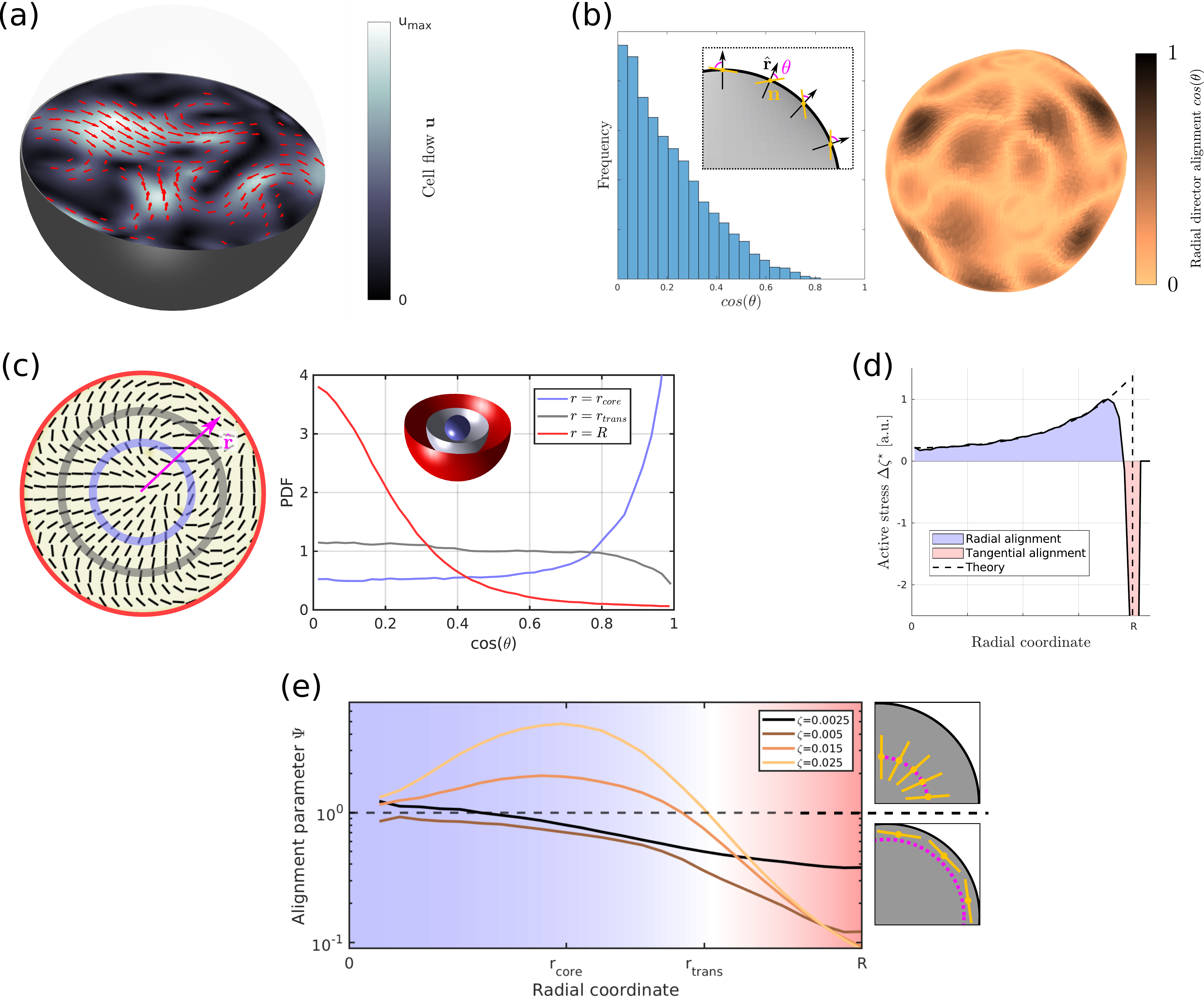}
	\caption{\textbf{(a)} Two-dimensional cross-section of the velocity field inside an active spheroid showing spatiotemporally chaotic flows caused by self-propelled disclination lines. Arrows and colour bar refer to the in-plane component of the flow field. \textbf{(b)} Histogram (left) and spatial distribution (right) of the angle $\theta$ between the director $\mathbf{n}$ and the radial vector $\hat{\mathbf{r}}$ at the surface of spheroids shows strong tangential director alignment caused by a hydrodynamic anchoring force termed active anchoring. \textbf{(c)} Two-dimensional cross-section of the in-plane component of the director field (left) and the probability distribution of $\cos \theta$ (right) obtained by spatial averages at a given radius $[r, r+\delta r]$ and over time. The distributions at radii $r_\text{core} < r_\text{trans} < R$ are shown as blue, grey and red, respectively. \textbf{(d)} $\Delta \zeta^\star$ obtained from simulations of quasi-spherical droplets. Activity gradients set up flows which create radial cell alignment inside the aggregate ($\Delta \zeta^\star > 0$, blue area) and tangential alignment close to the surface ($\Delta \zeta^\star < 0$, red area). The black, dashed line shows the analytical prediction eqn.~(\ref{eq:lapz}) for spherical aggregates with a sharp interface at $r=R$. \textbf{(e)} Cell alignment parameter $\Psi$ as a function of radial coordinate for different magnitudes of active stress $\zeta$. Simulations were performed using the following parameters, unless stated otherwise in the legend:} spheroid radius $R=30$, $\ell_m = 4$, $m_c=0.3$, $K_{LC}=0.04$, $\zeta=0.025$, $\xi=0$, $k_p=0$.
	\label{fig2}
\end{figure*}

We now consider the limit where cell flows are dominated by dipolar active stress $\zeta$ produced by cell division and death and growth-induced radial flows are negligible, $\mathbf{\nabla} \cdot \mathbf{u} \approx 0$. Active nematic stress creates a well-known hydrodynamic instability which constantly pushes the system out of equilibrium, leading to a state termed active turbulence. 3D active turbulence is characterised by spatiotemporally chaotic flows (Fig.~\ref{fig2} a) and the presence of self-propelled disclination lines (Fig.~\ref{figS3} a,b). In this section we will focus on how activity gradients affect the time-averaged alignment of the director at certain distances from spheroid centre, which is a quantity that can be easily measured in experiments \citep{desmaison2018impact, delarue2014stress}. 

To investigate how activity gradients affect cell alignment inside spheroids, we measure the distribution of the angle $\theta$ between the director $\mathbf{n}$ and the radial vector $\hat{\mathbf{r}}$ at different distances from the centre of aggregates. Since there are more possible configurations for angular alignment ($\theta=\pi/2$) than for radial alignment ($\theta=0$), a randomly aligned director field results in a uniform distribution of $\cos{\theta} \sim \mathcal{U}[0,1]$. Preferred angular or radial alignment leads to a biased distribution towards $0$ or $1$, respectively.  

Fig.~\ref{fig2} c shows a typical alignment profile across the spheroid.  
In the core cells are preferentially extended radially while close to the surface cells are extended along angular directions. In the transition region $r=r_{\text{trans}}$ between the shell and the core the director field has a random orientation.

Recalling that the active stress is $ \Pi_{act} = - \zeta^\star \mathbf{Q} $, (eqn.~(\ref{eq:activestress})), we can identify two competing mechanisms that lead to cell alignment.
Firstly, radial alignment in the bulk of spheroids is driven by gradients in activity, and hence in active stress, resulting from metabolite gradients. As was shown for bulk active nematic systems \citep{ruske2022activity}, hydrodynamic forces set up by activity gradients align the director field parallel to the gradient direction in regions where $\Delta \zeta^\star>0$ and perpendicular where $\Delta \zeta^\star<0$. Inside spherical aggregates eqns.~(\ref{eq:zetarad}), (\ref{eq:msphere}) yield
\begin{equation}
     \Delta \zeta^\star = \frac{\zeta}{\ell_m^2} \left[ \frac{R~\sinh \Tilde{r}}{r~\sinh \Tilde{R}} \right] \: ,
     \label{eq:lapz}
\end{equation}
which is positive throughout the cell aggregate, thus driving radial cell alignment.

Secondly, since the nematic droplet is surrounded by a passive isotropic fluid, the magnitude of active stress $\zeta^\star$ drops rapidly to zero across the surface of the aggregate. The resulting gradient in extensile activity at the surface  induces flows which rotate the director field and align it parallel to the surface as indicated by the distribution of the angle $\theta$ between the director field $\mathbf{n}$ and the surface normal of droplets in Fig.~\ref{fig2} b. This hydrodynamic effect, termed {\it active anchoring} \citep{blow2014biphasic,ruske2021morphology}, is restricted to the outer shell of spheroids, with a thickness given by the length-scale over which cell density $\varphi$ drops to zero, which is typically of the order of a single cell size (Fig.~\ref{fig2} d).

Thus the cell alignment profile in Fig.~\ref{fig2} c results from a competition between radial alignment due to activity gradients in the bulk, tangential alignment resulting from active anchoring at the aggregate surface and elastic forces which tend to smooth out distortions in the director profile.  To describe how the average cell alignment varies across aggregates with different magnitudes of active stress $\zeta$, we define the cell alignment parameter  
\begin{equation}
     \Psi = \frac{p(\cos \theta > 0.5)}{p(\cos \theta < 0.5)} \: ,
\end{equation}
which quantifies the degree of radial ($\Psi>1$) or angular ($\Psi<1$) cell alignment as a function of distance $r$ from the centre of spheroids (Fig.~\ref{fig2} e). 

If activity is sufficiently large ($\zeta \geq 0.015$) the director field shows strong radial alignment close to the core $r=r_{\text{core}}$ and tangential alignment close to the surface $r=R$, with weak or no alignment in the transition region $r=r_{\text{trans}}$. Note that, because of symmetry, there is no radial alignment at the centre of aggregates $r \approx 0$ since activity driven alignment cannot overcome the increasing elastic energy associated with a nematic hedgehog configuration close to the centre. 
These values of activity correspond to active turbulence and disclination lines are spontaneously created inside the aggregate (see also Fig.~\ref{figS3}).

At very low activity ($\zeta \leq 0.005$) we find that active anchoring creates weak tangential cell alignment close to the surface, with a magnitude that decreases with decreasing activity. The  angular alignment decays monotonically towards the core of spheroids without showing any radial alignment at intermediate distances because active stress is not strong enough to overcome elastic forces.

\subsection{\label{subsec:flowact} Interplay between proliferation- and activity-induced nematic alignment}

 \begin{figure*}[!ht]
	\centering
	\includegraphics[width=18cm]{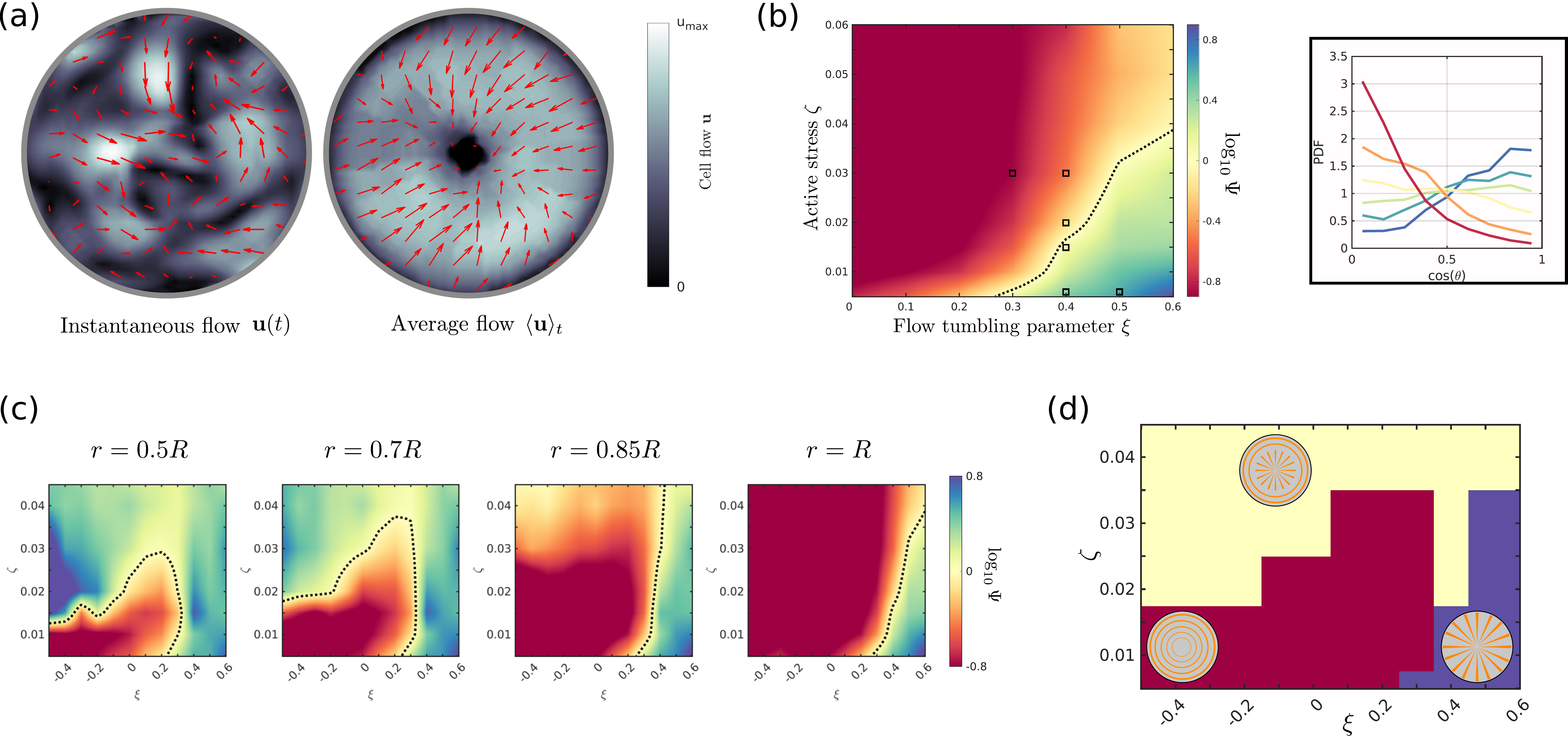}
	\caption{\textbf{(a)} Cross-section of an instantaneous velocity field (left) and time-averaged velocity field (right) inside spheroids with $k_p > 0$ and $\zeta > 0$. The chaotic flows of active turbulence seen in the instantaneous flow $\mathbf{u}$ average out over time-scales $T \gg \tau_a$, to give a a time-averaged flow $\langle \mathbf{u} \rangle_t$ resembling the converging flow-component driven by $\mathbf{\nabla} \cdot \mathbf{u} = k_p^\star$. Arrows and colour bar show the components of the flow field in the equatorial plane. The average velocity field $\langle \mathbf{u} \rangle_t$ is obtained by averaging over time $[0,1500\tau_a]$. \textbf{(b)} Heat map of the cell alignment parameter $\Psi$ at the surface of steady-state spheroids $r=R$ as a function of active stress magnitude $\zeta$ and flow-alignment parameter $\xi$. Red/blue regions correspond to tangential/radial surface alignment, respectively. The black, dotted line indicates the transition region where the director field has a random orientation ($\Psi=1$) because the flow-induced alignment balances active anchoring. The right panel shows the distribution of $\cos \theta$ from which $\Psi$ was calculated for six different parameter values indicated by black squares in the left panel. \textbf{(c)} Alignment parameter $\Psi$ shown at four different distances from the spheroid centre as a function of $\zeta$ and $\xi$. The shape of the contour line $\Psi=1$ (black dots) varies along the radial direction, converging to the distribution shown in panel (b) close the surface. \textbf{(d)} Three distinct regimes can be identified from variations of the cell alignment parameter at the surface, $\Psi_s=\Psi(R)$, and in the core, $\Psi_c = \Psi(R/2)$: homogeneous spheroids with radial (blue, $\Psi_c>0$, $\Psi_s>0$) or tangential orientation (red, $\Psi_c<0$  $\Psi_s<0$) and heterogeneous spheroids with tangential cell orientation close to the surface and radial alignment in the core (yellow, $\Psi_c>0$, $\Psi_s<0$).}
	\label{fig3}
\end{figure*}

We now combine the main findings of subsections ~\ref{subsec:flows} and \ref{subsec:active} and study the interplay between converging flows $\mathbf{\nabla} \cdot \mathbf{u} = k_p^\star$ set up by differential volume growth across spheroids and spatiotemporally chaotic flows created by dipolar stresses $\Pi_{act} = - \zeta^\star \mathbf{Q}$ arising from cell divisions and death. The spatial and temporal variations of flow fields in active turbulence can be characterized by an active length-scale $\ell_a = \sqrt{K_{LC}/\zeta}$ and time-scale $\tau_a = \eta/\zeta$, where $K_{LC}$ is the elastic constant of the liquid crystal which penalizes distortions in the director field resulting from cell shape deformations, and $\eta$ is the viscosity of the fluid. While instantaneous flow fields $\mathbf{u} (\mathbf{x},t)$ inside spheroids can be indistinguishable from active turbulence if active stress is sufficiently large, $\zeta/\eta > k_p$, chaotic flow patterns average out over time scales $T \gg \tau_a$ and the time-averaged velocity field $\langle \mathbf{u} (r)\rangle_t$ reduces to the converging cell flows given by eqn.~(\ref{eq:ur}) (Fig.~\ref{fig3} a).\\

We first consider the cell orientation at the surface of a spheroid, where the strain rate of converging flows reaches a maximum. Thus, following eqn.~(\ref{eq:Qrr}), the director orientation is most strongly affected by converging flows at the surface. Strains due to the converging component $\langle \mathbf{u} \rangle_t$ of the flow field, which is set up by eqn.~(\ref{eq:cont}), drive radial surface cell orientation for $\xi>0$, while active anchoring, driven by active stress $\Pi_{act}$ in eqn.~(\ref{eq:NSE}), favours cells aligning tangentially to the surface. The magnitudes of proliferation-driven alignment and active anchoring increase with the flow-alignment parameter $|\xi|$ and active stress $\zeta$, respectively. To investigate the cross-over between these competing effects, we measure the cell alignment parameter $\Psi$ at the surface of steady-state spheroids over a range of parameter values $\xi$ and $\zeta$ (Fig.~\ref{fig3} b). For $\xi = 0$ the orientation of cells does not respond to the strain rate of flows, thus active anchoring dominates at any activity level $\zeta>0$, which creates strong tangential surface alignment. As $\xi>0$ increases, however, the cell orientation responds more strongly to the fluid strain rate at the surface and the director field aligns radially due to velocity gradients caused by proliferation as long as active anchoring is sufficiently small, $\zeta < \zeta_c$. Since shear-driven cell alignment scales as $\dot{Q}_{ij} \sim \xi \Tilde{E}_{ij}$, the critical activity $\zeta_c$ above which active anchoring dominates increases with the strain rate at the surface $\Tilde{E}_{rr} \sim k_p$ and $\xi$. The function $\zeta_c(\xi)$ is shown by the black, dotted line in Fig.~\ref{fig3}~b. 

In addition to cell orientation at the surface, we can also measure $\Psi$ at a distance $0<r<R$ from the centre of spheroids as a function of $\zeta$ and $\xi$ (Fig.~\ref{fig3} c). For low activity and $\xi=0$, the director field in the core ($r=0.5 R$) shows weak alignment since the orientation at the surface decays very slowly towards the centre due to the dominance of the elastic energy associated with distortions in the director field over active stresses. Thus, in the regime of low activity, cell orientation throughout spheroids is  dictated by the cell alignment at the surface. As soon as radial shear-alignment becomes dominant at the surface for $\xi>\xi_c$, cells globally switch from angular to radial orientation (see phase diagram Fig.~\ref{fig3} d for $\zeta<0.02$).  

In section \ref{subsec:active} we showed that for sufficiently large activity $\zeta>\zeta_c$, the core of spheroids shows radial alignment while at the surface cells are oriented tangentially because of the balance between activity gradient-induced radial ordering and tangential active anchoring. Surprisingly, the activity threshold $\zeta_c$ above which radial order is created in the core decreases for $\xi<0$, even though converging flows favour tangential over radial alignment throughout spheroids for $\xi<0$ (see boundary between red and yellow domains in Fig.~\ref{fig3} d). This non-trivial behaviour can be explained in terms of the disclination line structure inside aggregates. Active anchoring creates significant surface alignment, but usually it is not strong enough to completely enforce tangential anchoring everywhere on the surface (see Fig.~\ref{fig2} b). As $\xi<0$ decreases, however, the tangential surface anchoring of the director field eventually becomes sufficiently large that the Gauss–Bonnet theorem can be applied. This states that the total topological charge of a vector field tangentially anchored on a sphere must be $+2$, the Euler-characteristic $\chi$ of the sphere \citep{eisenberg1979proof}. 

At very low activity and $\xi<0$, the director field thus relaxes into a configuration that minimizes its elastic energy while meeting the topological constraint close to the surface. The resulting defect structure is two $+1/2$ disclination lines which bend towards the centre of the droplet and create four $+1/2$ defects at the surface. The cross-section of the director field inside droplets is formed by two $+1/2$ defects located close to the centre and hence resembles that around a two-dimensional $+1$ defect  (Fig.~\ref{figS4} a,d). Due to the large elastic energy associated with a point-like $+1$ defect, there is a finite repulsion force between the two $+1/2$ defects which keeps them at a finite distance. On the other hand, contractile active forces on $+1/2$ defects oppose the elastic forces, thus the distance between the two stationary disclination lines decreases with increasing activity.

If activity reaches a critical threshold $\zeta_c(\xi)$, the two $+1/2$ defects at the centre slip past each other and subsequently reach a steady-state in which both defects orbit around the centre, thereby creating persistent rotational motion in the core of spheroids (Fig.~\ref{figS4} b). In 3D this corresponds to a \textit{dancing disclination} state, where the $+1/2$ line segments in the core orbit around the centre, while the end-points of disclination lines at the surface move very slowly. This motion makes it inevitable that disclination lines must cross after each full rotation, thereby rewiring some of the line segments in the core (Fig.~\ref{figS4} e). We note that the rotational motion of disclination lines is reminiscent of persistent clockwise or counter-clockwise rotations in 2D active nematic systems in circular confinement which has been observed in experiments \citep{opathalage2019self} and simulations \citep{gao2017analytical, young2021many}. The orientation of $+1/2$ defects in the dancing disclination state creates significant radial alignment of the director field outside the orbiting defects, hence $\Psi$ shows a sharp transition from angular to radial director alignment in the core at $\zeta=\zeta_c(\xi)$ (Fig.~\ref{fig3} c). As the activity $\zeta$ is increased even further, the motion of the disclination lines becomes more chaotic and eventually the oscillations are lost in the active turbulent background (Fig.~\ref{figS4} c,f).

\subsection{\label{subsec:infer} Parameter inference for biological systems}

\begin{figure*}
	\centering
	\includegraphics[width=18cm]{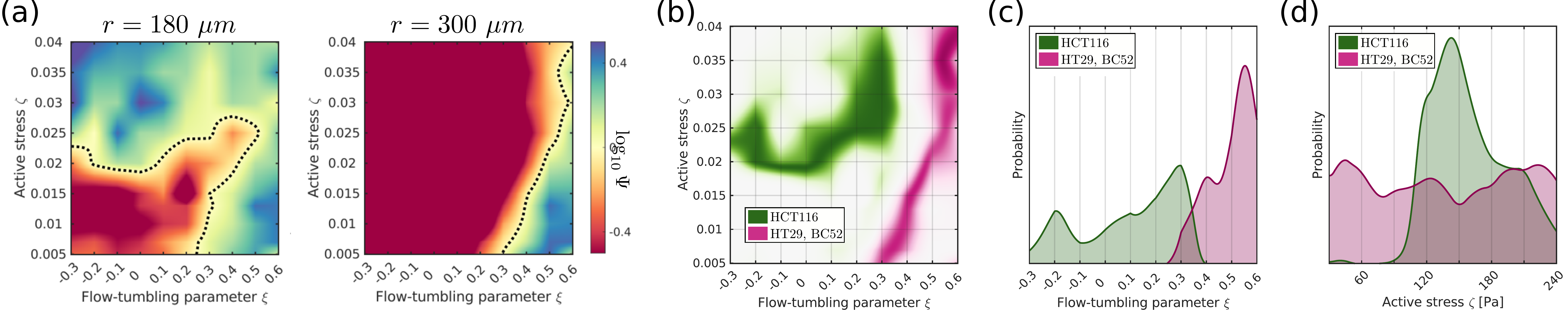}
	\caption{\textbf{(a)} Cell alignment $\Psi$ in the core and close to the surface of spheroids of size $R=300 \mu m$ using simulation parameters estimated from experimental data (see appendix \ref{appx:parmap}). \textbf{(b)} Posterior distributions $p(\zeta,\xi|\mathbf{X})$ of activity $\zeta$ and flow-alignment parameter $\xi$ inferred from cell orientations inside HCT116 (red) and  HT29/BC52 spheroids (blue) using $\epsilon = 0.05$. Based on the available data \citep{desmaison2018impact, delarue2014stress}, we estimated $\mathbf{X}=[\log \Psi^{exp}_c>0.1, \log \Psi^{exp}_s=0]$  for HT29/BC52 spheroids and $\mathbf{X} = [\log \Psi^{exp}_c=0, \log \Psi^{exp}_s<-0.4]$ for HCT116. \textbf{(c,d)} Marginal distributions $p(\xi|\mathbf{X})$ and $p(\zeta|\mathbf{X})$.}
	\label{fig4}
\end{figure*}

Finally we apply our theoretical active fluid model to experimental data obtained from freely grown proliferating spheroids to make quantitative predictions about dynamical tissue parameters, such as active stress $\zeta$ and the flow-alignment parameter $\xi$. Since it is rarely possible in experiments to simultaneously measure proliferation gradients, active stress, cell flows and cell orientations inside spheroids, we require physical estimates for model parameters $\ell_m$, $m_c$ and $k_p$.  
 
 The characteristic penetration length $\ell_m$ of metabolites can be estimated by measuring proliferation profiles in spheroids. This can be realized either by using immuno-fluorescence staining procedures to mark proliferating cells \citep{yerushalmi2010ki67} or by using spheroid models expressing Fucci (Fluorescence Ubiquitination Cell Cycle Indicators) tools that allow monitoring the cell cycle position of a cell \citep{sakaue2008visualizing}. The metabolite penetration length from the surface can vary upon the cell packing density and the cell type, however, for most types of human tumor cells grown under optimal nutrient and oxygen conditions the penetration length typically varies between $\ell_m \approx 100 - 200~\mu m$ \citep{sutherland1988cell, laurent2013multicellular, larue2004microenvironmental}. Assuming spheroids are at steady-state, the critical metabolite concentration $m_c$ can then be inferred from the aggregate size $R_c$ using eqn.~(\ref{eq:Rdot}), $m_c=3(x \coth x - 1)/x^2$ , where $x=R_c/\ell_m$ (Fig.~\ref{figS1} a). 

The cell division rate $k_+$ is determined by the time interval between cell divisions, the cell-cycle time $T_C$, and the proportion of cells engaged in the cell cycle, the growth fraction $\mathfrak{f}$, such that $k_+ = \mathfrak{f}/T_C$. The growth fraction $\mathfrak{f}(\mathbf{x})$ depends on the local metabolite concentration and varies within spheroids \citep{laurent2013multicellular, montel2012isotropic}. Assuming $k_p^\star = k_+$ at the surface (no cell death), the growth magnitude $k_p$ can be most easily estimated from the growth fraction at the surface of spheroids,
\begin{equation}
    k_p = \mathfrak{f}^{s}/[T_c(1-m_c)] \: ,
\end{equation}
where $\mathfrak{f}^{s}$ typically varies between $0.6 - 0.9$ \citep{alessandri2013cellular, montel2012isotropic, laurent2013multicellular,lee2019dispersible}. The cell-cycle time in carcinoma spheroids and tumors is typically around $T_C \approx 20~h$, with slight variations across different cell types \citep{eidukevicius2005method, delarue2013mechanical}.\\

To validate the proposed parameter estimates, we compare our model to a system of freely grown CT26 spheroids at steady-state with radius $R_c=450~\mu m$, corresponding to $m_c \approx 0.7$. In the outer layer of the spheroids, $0.8 < r/R < 1$, eqn.~(\ref{eq:ur}) predicts an average radial flow velocity $u_r \approx 0.06 ~ \mathfrak{f}^{s}/T_c \approx 26~\mu m~\text{day}^{-1}$, which is in good agreement with the radial flow measured in experiments using fluorescently labeled particles, $u_r^{exp} = 21~\mu m~\text{day}^{-1}$ \citep{delarue2013mechanical}.

Mitotic cells in three-dimensional microenvironments generate substantial protrusive forces which were estimated to be on the order of $\zeta_+ \approx 1 - 2~kPa$ \citep{nam2018mitotic}. Considering that the mitotic time $T_M$ over which strong anisotropic forces are generated is relatively short compared to the cell cycle time, $T_M/T_C \approx 0.1$, the average stress produced by cell divisions over a cell cycle is $\zeta \approx 100 - 200~Pa$. Similar magnitudes of cell-generated anisotropic stress were measured within proliferating multicellular spheroids using fluorescently labelled, cell-sized oil droplets or polyacrylamide microspheres to measure stresses in situ \citep{lucio2017spatiotemporal,lee2019dispersible}. 

We now compare our active fluid model to experimental data obtained from freely grown proliferating spheroids ($R \approx 300 \mu m$) consisting of human colon carcinoma cells (HCT116) \citep{desmaison2018impact},  colon adenocarcinoma cell (HT29) and human breast cancer cells (BC52) \citep{delarue2014stress}. It has been reported that both the orientation and the division axis of cells inside HCT116 spheroids show significant tangential alignment close to the surface ($\Psi_s^{exp}<1$, $cos \langle \theta \rangle = 0.22$) while showing no significant alignment in the core ($\Psi^{exp}_c \approx 1$, $cos \langle \theta \rangle = 0.44$). On the other hand, the analysis of cell shapes and polarity inside HT29 and BC52 spheroids revealed that cells show only weak surface alignment ($\Psi^{exp}_s \approx 1$) while cells are increasingly radial elongated towards the core ($\Psi^{exp}_c>1$).

Assuming $\ell_m=150~\mu m$, $\mathfrak{f}^{s}=0.7$ and $T_C = 20~h$, we calculated the average cell alignment $\Psi_s$ at the surface ($r=300 \mu m$) and $\Psi_c$ in the core of spheroids ($r=180 \mu m$) over a range of values for $\zeta$ and $\xi$ (Fig.~\ref{fig4} a). The full list of simulation parameters with their mapping from lattice units to physical values are shown in appendix \ref{appx:parmap}. By comparing $\Psi_s$ and $\Psi_c$ obtained from simulations with experimental data, we calculate the posterior distributions $p(\Theta|\mathbf{X})$ over model parameters $\Theta = [\zeta,\xi]$ given data $\mathbf{X} = [\Psi_s^{exp}, \Psi_c^{exp}]$. Given the qualitative nature of the experimental data, we assume that the likelihood function $\mathcal{L}(\mathbf{X}|\Theta)$ follows a normal distribution, $\log \Psi(\Theta) \sim \mathcal{N}(\mu=\log \Psi^{exp},\sigma=0.2)$. The parameter distributions $p(\zeta,\xi|\mathbf{X})$ obtained for HCT116 (green) and HT29/BC52 spheroids (magenta) are presented in Fig.~\ref{fig4} b. We find that tissue dynamics within HT29 and BC52 spheroids is governed by a positive flow-alignment parameter ($\xi \approx 0.5$) while in HCT116 spheroids shear-alignment of cells is less dominant ($\xi<0.3$) and potentially even negative (Fig.~\ref{fig4} c). We also gain some additional information about the magnitude of active stress inside aggregates, where cell divisions inside HCT116 spheroids produce on average $\zeta = 100 - 200 Pa$ (Fig.~\ref{fig4} c), thus confirming our previous experimental estimate based on $\zeta = \zeta_+ T_M / T_C$.

\section{\label{sec:conclusion} Conclusion}

We have investigated how radial flows and active stresses affect cell alignment inside proliferating multicellular spheroids using three-dimensional active nematic droplets as a model system. Cell orientation and flows inside aggregates are described by a continuous director $\mathbf{n}$ and velocity field $\mathbf{u}$, which follow the hydrodynamic equations of motion of active nematics. Diffusive transport of metabolites creates proliferation gradients across spheroids, where cells close to the surface have sufficient access to metabolites and divide, while metabolites are depleted in the core of spheroids and cells die. Since cell division/death is associated with a mass source/sink and extensile/contractile active stress along the cell orientation axis, proliferation gradients induce radial cell flows and activity gradients inside spheroids. 

We show that converging cell flows inside steady-state spheroids $can$ promote radial or circumferential cell alignment, depending on the sign of the flow-alignment parameter $\xi$, a tissue parameter quantifying how cell orientations respond to shear-flows. The magnitude of shear-driven alignment scales with $\xi$ and with the strain rate, the latter reaching a maximum at the surface and decreasing towards the core.

Active stress, on the other hand, creates a well-known hydrodynamic instability which leads to a chaotic steady state termed active turbulence. Activity gradients impact the average director field orientation inside spheroids in two ways: first, smooth variations between contractile stress in the core and extensile stress at the surface of spheroids drive radial cell alignment by creating flows which reorient the director field \citep{ruske2022activity}. The second contribution is active anchoring created by the sudden drop of extensile activity across the interface of spheroids, which are embedded in a passive fluid. The resulting sharp activity gradients at the surface induce flows which rotate the director field parallel to the interface, thus creating strong tangential surface alignment. Therefore gradients in active stress are responsible for radial core alignment and tangential surface alignment.

Depending on the relative strength of flow-driven shear-alignment and activity induced alignment, we find three distinct cell orientation regimes inside spheroids: for small activity, cell orientation is either tangential throughout the aggregate for $\xi<\xi_c$ or radial for $\xi>\xi_c>0$. If active stress becomes sufficiently large, $\zeta > \zeta_c(\xi)$, one recovers spheroids with radial core alignment and tangential surface alignment (Fig.~\ref{fig3} d).

The cell orientation profile inside aggregates significantly affects the stress distribution and short-time response of spheroids to external pressure jumps \citep{delarue2014stress}. It has also been shown that the elongation and orientation of cells within fibrosarcoma spheroids contribute to invasive behaviour by priming cells at the spheroid periphery to exhibit the correct morphology and polarisation for effective invasion into the surrounding matrix \citep{valencia2015collective}. The activity level and flow alignment properties of cellular aggregates could thus be used as control parameters to guide the migratory behaviour of spheroids and their mechanical response to changes in their environment.

Furthermore, we apply our model to experimental systems of freely grown HCT116, HT29 and BC52 spheroids to infer distributions of tissue parameters $\xi$ and $\zeta$ based on the distribution of cell orientations inside spheroids. We find a significant difference in shear-alignment across different systems, with $\xi \approx 0.5$ for HT29/BC52 spheroids, while for HCT116 spheroids shear-alignment $\xi<0.3$ is less dominant and potentially even negative. It is of interest to compare these results with previous measurements of cell motion in two-dimensional systems, which indicated that cell orientation and flows in dense monolayers of myoblasts and epithelial cells are governed by a negative flow-alignment parameter $\nu$, which corresponds to $\xi>0$ \citep{aigouy2010cell, duclos2018spontaneous}. This suggests that the same biomechanical mechanisms that cause flow-induced realignment of cells in monolayers may play an important role in three-dimensional tissue organisation.

In this paper we have focused on tissue dynamics in the hydrodynamic limit, where cell aggregates are described as a viscous fluid on long time scales. This continuum model could be easily generalized to viscoelastic spheroids by also including an elastic response on time scales smaller than a characteristic relaxation time $\tau$. The viscoelastic relaxation time of cell aggregates, which is typically on the order of $\tau \approx 5-40~\mbox{min}$ \citep{guevorkian2010aspiration, marmottant2009role}, is much smaller than typical cell-cycle times of $T_C\approx 20~\mbox{h}$, thus elastic contributions should be negligible on the time scales of freely grown spheroids. However, cell compressibility and the elastic properties of spheroids become important when they are subject to fast environmental changes, such as external pressure jumps \citep{delarue2014stress,delarue2014compressive, alessandri2013cellular}. Mechanical confinement or compressive stress also alters proliferation profiles inside spheroids, where $k_+$ decreases with increasing pressure, while $k_-$ stays approximately constant \citep{desmaison2013mechanical, montel2012isotropic}. Our model could also be used to study systems in which the surface tension of spheroids is sufficiently small that active stress can strongly deform spherical aggregates. This would allow avascular spheroids to overcome diffusion-limited, spherical growth by forming protrusions, thereby creating a branched network structure that grows indefinitely in environments with an unlimited supply of nutrients.

We hope that recent developments in experimental techniques and data processing will permit the measurement of 3D cell orientations inside aggregates with high spatial and temporal resolution in the future. This would allow the inference of more complex dynamics, including features like non-linear relationships between proliferation rate $k_p^\star$ and metabolite concentration $m$ or spatial variations of $\xi$ within aggregates. Furthermore, data with a sufficient temporal resolution would allow the testing of hypotheses relating to the underlying dynamics of cells, such as a possible dependence of intrinsic activity $\zeta_0$ or flow-alignment $\xi$ on internal cell states, giving rise to time-dependent parameters $\zeta^\star (t)$ and $\xi(t)$, which could be inferred from the cell alignment profiles. Here we highlight the potential of active fluid theories to model complex biological systems and show that cell orientation profiles in proliferating spheroids can be used to extract dynamical tissue parameters, which are otherwise difficult to measure directly.

\section*{Acknowledgements}
This project was funded by the European Commission’s Horizon 2020 research and innovation programme under the Marie Sklodowska-Curie grant agreement No 812780.

\FloatBarrier
\appendix

\section{\label{appx:eom} Expressions for the free energy and hydrodynamic equations of motion}

Rod-like particles can not only be advected by the fluid, but also rotate in response to flow gradients. This behaviour is accounted for by the co-rotational term \citep{beris1994thermodynamics} %
\begin{multline}
	\mathcal{W}_{ij} = \left( \xi \Tilde{E}_{ik}+\Omega_{ik} \right) \left( Q_{kj} + \frac{\delta_{kj}}{3} \right) + \left( Q_{ik} + \frac{\delta_{ik}}{3} \right) \\
	\left( \xi \Tilde{E}_{kj}-\Omega_{kj} \right) - 2\xi \left( Q_{ij} + \frac{\delta_{ij}}{3} \right) Q_{kl} W_{lk} \: ,
	\label{eq:corot}
\end{multline}
where $\Tilde{E}_{ij} = E_{ij} - \delta_{ij} E_{kk}/3 $ is the traceless part of the strain rate tensor $E_{ij}=(\partial_j u_i + \partial_i u_j)/2$ and
$\Omega_{ij}=(\partial_j u_i - \partial_i u_j)/2$ is the antisymmetric part of the velocity gradient tensor $W_{ij}=\partial_i u_j$. The alignment parameter $\xi$ quantifies how the director responds to pure shear flow.

The passive contributions of the hydrodynamic stress tensor $\Pi_{passive}$ in the Navier-Stokes equations are
\begin{equation}
 \Pi_{viscous} = 2 \eta \mathbf{D} \:,
\end{equation}
 \begin{multline}
\Pi_{capillary} = (f-\mu \varphi) \mathbf{I} - \mathbf{\nabla} \varphi \left( \frac{\partial f}{\partial (\mathbf{\nabla} \varphi)} \right) \\
 +\mathbf{\nabla} \varphi \nabla \left( \frac{\partial f}{\partial (\mathbf{\nabla}^2 \varphi)} \right) - \mathbf{\nabla} \mathbf{\nabla} \varphi \left( \frac{\partial f}{\partial (\mathbf{\nabla}^2 \varphi)} \right)\: ,
 \end{multline}
\begin{multline}
 \Pi_{elastic} = -p \mathbf{I} - \xi [ \mathbf{H} \Tilde{\mathbf{Q}} + \Tilde{\mathbf{Q}} \mathbf{H} - 2 \Tilde{\mathbf{Q}}  tr(\mathbf{Q} \mathbf{H}) ] \\
 + \mathbf{Q} \mathbf{H} - \mathbf{H} \mathbf{Q} - \mathbf{\nabla} \mathbf{Q} \left( \frac{\partial f}{\partial (\mathbf{\nabla} \mathbf{Q})} \right) \: , \\
 \end{multline}
where $\rho$ is the density, $\eta$ the viscosity, $p$ the bulk pressure and $\Tilde{\mathbf{Q}} = \left(\mathbf{Q}+\frac{1}{3} \mathbf{I}\right)$. The free energy density $f=f_{LC}+f_{GL}$ consists of a liquid crystal component $f_{LC}$ for the orientational order parameter $\mathbf{Q}$ and a Ginzburg-Landau contribution $f_{GL}$ for a concentration field $\varphi$ which is described below. 

The mechanical and geometric properties of cells are accounted for by choosing an appropriate nematic free energy density of the system
\begin{multline}
	f_{LC} = A_{LC} \Bigg\{ \frac{1}{2} \left( 1-\frac{\Bar{\eta}(\varphi)}{3}\right) \:  \mbox{tr}(\mathbf{Q}^{2}) - \frac{\Bar{\eta}(\varphi)}{3} \:  \mbox{tr}(\mathbf{Q}^{3})  \\
	+ \frac{\Bar{\eta}(\varphi)}{4} \: \mbox{tr}(\mathbf{Q}^{2})^{2}  \Bigg\} + \frac{1}{2} K_{LC} \left( \mathbf{\nabla} \mathbf{Q} \right)^{2} \: 
	\label{eq:LCbulk}
\end{multline}
which includes the usual Landau-de Gennes bulk energy of the liquid crystal and a term which penalizes elastic deformations of the director field \citep{de1993physics}.


We follow the shape of the growing spheroids by modelling them as deformable, nematic droplets in an isotropic fluid background. This is achieved by solving the reaction-diffusion equation of a concentration field $\varphi(\mathbf{x},t)$, as described in detail in previous work \citep{ruske2021morphology}:
\begin{equation}
	\left( \partial_{t} + \mathbf{u} \cdot \mathbf{\nabla} \right) \varphi = \Gamma_{\varphi} \: \mathbf{\nabla}^{2} \mu\:.
	\label{EoM_phi}
\end{equation}
Here $\mathbf{u}$ is the velocity field and the mobility $\Gamma_{\varphi}$ quantifies how fast $\varphi$ responds to gradients in the chemical potential $\mu=\delta \mathcal{f}/\delta \varphi$. The free energy density $f_{GL}$ is chosen to take the Ginzburg-Landau form
\begin{equation}
	f_{GL} = A_\varphi \varphi^2 \left(1 -\varphi\right)^2 + \frac{K_{\varphi}}{2} \left( \mathbf{\nabla} \varphi \right)^{2} \: .
\end{equation}
This describes phase separation into two stable phases with concentrations $\varphi=0,1$ and with an interface of width $L\sim\sqrt{A_\varphi/K_\varphi}$ which separates the inside ($\varphi=1$) and outside ($\varphi=0$) of spheriods and introduces a surface tension $\gamma \sim \sqrt{A_{\varphi} K_\varphi}$ \citep{cahn1958free}. Parameters $A_{\varphi}$ and $K_{\varphi}$ are chosen to match the surface tension $\gamma$ measured in experiments, while ensuring that the interface width is much smaller than any other length-scale in the system, $L \ll \text{min}[\ell_a,\ell_m]$. Since cell aggregates are modelled using a continuous concentration field $0 \leq \varphi \leq 1$, the net cell production rate and active stress in simulations are continuous functions and follow $k_p^{LB}=\varphi~k_p$ and $\zeta^{LB}=\varphi~\zeta$, respectively (see Fig.~\ref{figS2}).

Throughout the paper we use the following simulation parameters, in lattice-units, unless otherwise stated: $\rho=1$, $\eta = 2/3$, $A_\varphi = 0.2$, $K_\varphi = 0.4$, $\Gamma_\varphi=0.2$, $A_{LC} = 1.5$, $K_{LC} = 0.03$, $\Gamma=0.1$. To highlight the effects of flow-induced elongation of isotropic cells in section \ref{subsec:flows}, we chose $\Bar{\eta} = 2.55 + 0.1 \varphi$ which creates an isotropic phase $S_{eq}=0$ inside droplets unless flows drive nematic order for $\xi \neq 0$. In sections \ref{subsec:active}, \ref{subsec:flowact}, \ref{subsec:infer}, we chose $\Bar{\eta} = 2.7 + 0.3(\varphi-0.5)$ which creates nematic order $S_{eq}=0.3$ of constant magnitude inside aggregates ($\varphi=1$), while ensuring that the fluid environment stays isotropic ($\varphi=0$).

\section{\label{appx:stbanly} Stability analysis of isotropic spheroids}

We derive the stability of the isotropic state $Q_{ij} = 0 + Q^{'}_{ij}$ by expanding the Beris-Edwards eqn.~(\ref{eq:Qt}) to first order in perturbation parameter $Q^{'}_{ij} = \epsilon (n_i n_j - \delta_{ij}/3)$,
\begin{equation}
    D_t Q^{'}_{ij} - \mathcal{W}_{ij} = \Gamma K_{LC} \left( \partial_k \partial_k Q^{'}_{ij} \right) + \mathcal{O}(\epsilon^2)\: .
\end{equation}
Following eqn.~(\ref{eq:corot}), the leading order contribution of the co-rotation term $\mathcal{W}_{ij}$ is 
\begin{equation}
    \mathcal{W}_{ij} = \frac{2}{3} \xi \Tilde{E}_{ij} + \mathcal{O}(\epsilon)\: ,
\end{equation}
where we have used the symmetry of $\Tilde{E}_{ij}$ and $Q^{'}_{ij}$. The stability of the isotropic state is thus governed by
\begin{equation}
    \dot{Q}_{ij} = \frac{2}{3} \xi \left( E_{ij} - \delta_{ij} \frac{E_{kk}}{3} \right) + \mathcal{O}(\epsilon)\: ,
\end{equation}
where we have neglected the advective term, $u_k \partial_k Q_{ij}$, since spatial gradients $\mathbf{\nabla} \mathbf{Q}$ vanish in the isotropic state. In spherical coordinates the strain rate tensor $E_{ij}=(\partial_j u_i + \partial_i u_j)/2$ has components
\begin{align}
    E_{rr} &= \partial_r u_r \: , \\
    E_{\theta \theta} &= E_{\phi \phi} = \frac{u_r}{r} \: , \\
    E_{r\theta} &= E_{r\phi} = E_{\theta \phi} = 0 \: ,
\end{align}
where we have used the symmetry of the flow field $\mathbf{u}(r,\theta,\phi) = u_r(r) \hat{\mathbf{r}}$. This yields the initial growth rate of the radial component of the nematic order
\begin{equation}
    \dot{Q}_{rr} = \frac{4}{9} \xi \left( \partial_r u_r - \frac{u_r}{r} \right) \: .
\end{equation}

\section{\label{appx:parmap} Mapping of simulation parameters to experimental values}

In order to map lattice Boltzmann (LB) simulation parameters to dimensional quantities in physical units, one requires a physical reference scale for three independent LB parameters, such as the lattice spacing $\delta x$, a force scale and the viscosity. For modelling the organisation of three-dimensional, multicellular spheroids, we chose a lattice spacing such that the total diameter of an aggregate is about 50 lattice sites, $\delta x \approx 0.04 \: R$. To estimate the bending rigidity of cells $K_{LC}$ in experiments, we need to relate the mechanical properties of individual cells to the effective constant $K_{LC}$ of the nematic description. If we assume that  cells in a dense aggregate must physically deform when the nematic order is distorted, the elastic energy associated with cell shape deformations is related to the cells' Young modulus $E$ and cell size $L$. From dimensional arguments, it follows that $K_{LC} \sim E \cdot L^2$ \cite{duclos2014perfect}. The typical size and Young's modulus of colon and breast cancer cell lines is of the order of $L \sim 10 \: \mu m$ and $E \sim 100 \: Pa$ \cite{pachenari2014mechanical, li2008afm, nematbakhsh2017correlating}, respectively, which yields $K_{LC}\sim 10^{-8} N$. 
Choosing $K_{LC}= 5 \cdot 10^{-8} N$ and the apparent viscosity of cell aggregates $\eta \approx 60 \: kPa \: s$ \cite{forgacs1998viscoelastic, duclut2019fluid} as LB reference scales, the LB parameters map to the following physical units:

\FloatBarrier
\begin{table}[!ht]
  \begin{center}
    \label{LB_units}
    \begin{tabular}{l|c|r}
      \textbf{Parameter} & \textbf{LB units} & \textbf{Physical units}\\
      \hline
      Aggregate size $R$ & $25$ & $300 \: \mu m$\\
      Bending rigidity $K_{LC}$ & $0.03$ & $5 \cdot 10^{-8} \: N$\\
      Viscosity $\eta$ & $2/3$ & $60 \: kPa \: s$\\
      \hline
      Lattice spacing $\delta x$ & $1$ & $12 \: \mu m$\\
      Time step $\delta t$ & $1$ & $15 \:s$\\
      Critical concentration $m_c$ & $0.8$ & $0.8$\\
      Penetration length $\ell_M$ & $12$ & $150 \: \mu m$\\
      Growth rate $k_p$ & $8 \cdot 10^{-4}$ & $5 \cdot 10^{-5} \: s^{-1}$\\
      Active stress $\zeta$ & $0.005 - 0.04$  & $30 - 240 \: Pa$ \\
      \hline
      Surface tension $\gamma$ & $0.3$ & $  4 \cdot 10^{-2} \: N/m$\\ 
      Spheroid bulk energy $A_{\varphi}$ & $0.2$ & $2.4 \: kPa$\\
      Nematic bulk energy $A_{LC}$ & $1.5$ & $18 \: kPa$\\
      Rotational diffusivity $\Gamma$ & $0.1$ & $6 \cdot 10^{-7} \: (Pa \: s)^{-1}$\\
      Mobility $\Gamma_{\varphi}$ & $0.2$ & $8\cdot 10^{-16} \: m^2/Pa \: s$\\
    \end{tabular}
  \end{center}
\end{table}
\FloatBarrier

As outlined in section \ref{subsec:infer}, these parameters are in good agreement with mechanical properties of cell aggregates measured in experiments, with typical tissue interfacial tensions $\gamma \sim 10^{-2} \: N/m$ \cite{stirbat2013fine, forgacs1998viscoelastic, guevorkian2010aspiration, mgharbel2009measuring}. It should be noted, however, that the mechanical properties of cell aggregates may vary greatly as the Young's modulus varies over several orders of magnitude, $E \sim 0.1 - 10 \: kPa$ for different cell types \cite{blumlein2017mechanical, kuznetsova2007atomic}. Similar variations are observed for the effective viscosity of tissues and cell aggregates, $\eta \sim 10 - 100 \: kPa \: s$ \cite{guevorkian2010aspiration,marmottant2009role,stirbat2013fine}. 

\section{\label{appx:discl} Detection of disclination lines and calculation of the twist angle $\beta$}
We use Zapotoky's defect-finding algorithm \cite{zapotocky1995kinetics} to find defect positions on three-dimensional grids. This approach checks if a disclination is located at the intersection of four voxels forming a $2 \times 2$ square repeated along all three coordinate axis \cite{hobdell1997numerical}. When a disclination is found, the rotation vector $\Omega$ along which the director field winds is determined by taking the cross product of each pair of directors around it. Once all grid points are classified, continuous disclination lines are identified as the shortest line connecting all defect positions, and the twist-angle $\beta$ can be obtained by measuring the angle between $\Omega$ and the local line tangent. A shortcoming of this algorithm is that the rotation vector $\Omega$ has an arbitrary sign and therefore is unsuited to distinguishing between $\beta=0$ ($-1/2$-type) and $\beta=\pi$ ($+1/2$-type) disclinations.
To achieve this, we calculate the saddle-splay energy
\begin{equation}
    \Bar{f}_{24} = \mathbf{\nabla} \cdot \left[(\mathbf{n} \cdot \mathbf{\nabla})\mathbf{n} - \mathbf{n}(\mathbf{\nabla} \cdot \mathbf{n}) \right] \: ,
\end{equation}
which is negative for $\beta=\pi$ line segments, positive at $\beta=0$ segments and zero for twist defects \cite{tran2016lassoing}. The calculation of the line tangent is performed on a discrete grid, hence the local tangent does not vary continuously along disclination lines which leads to the small color jumps seen for some disclination line segments in Fig.~\ref{figS3} and Fig.~\ref{figS4}.

\onecolumngrid
\newpage
\FloatBarrier
\section{\label{appx:suppfig} Supplementary Figures}

\setcounter{figure}{0}
\renewcommand{\thefigure}{S\arabic{figure}}%

 \begin{figure*}
	\centering
	\includegraphics[width=18cm]{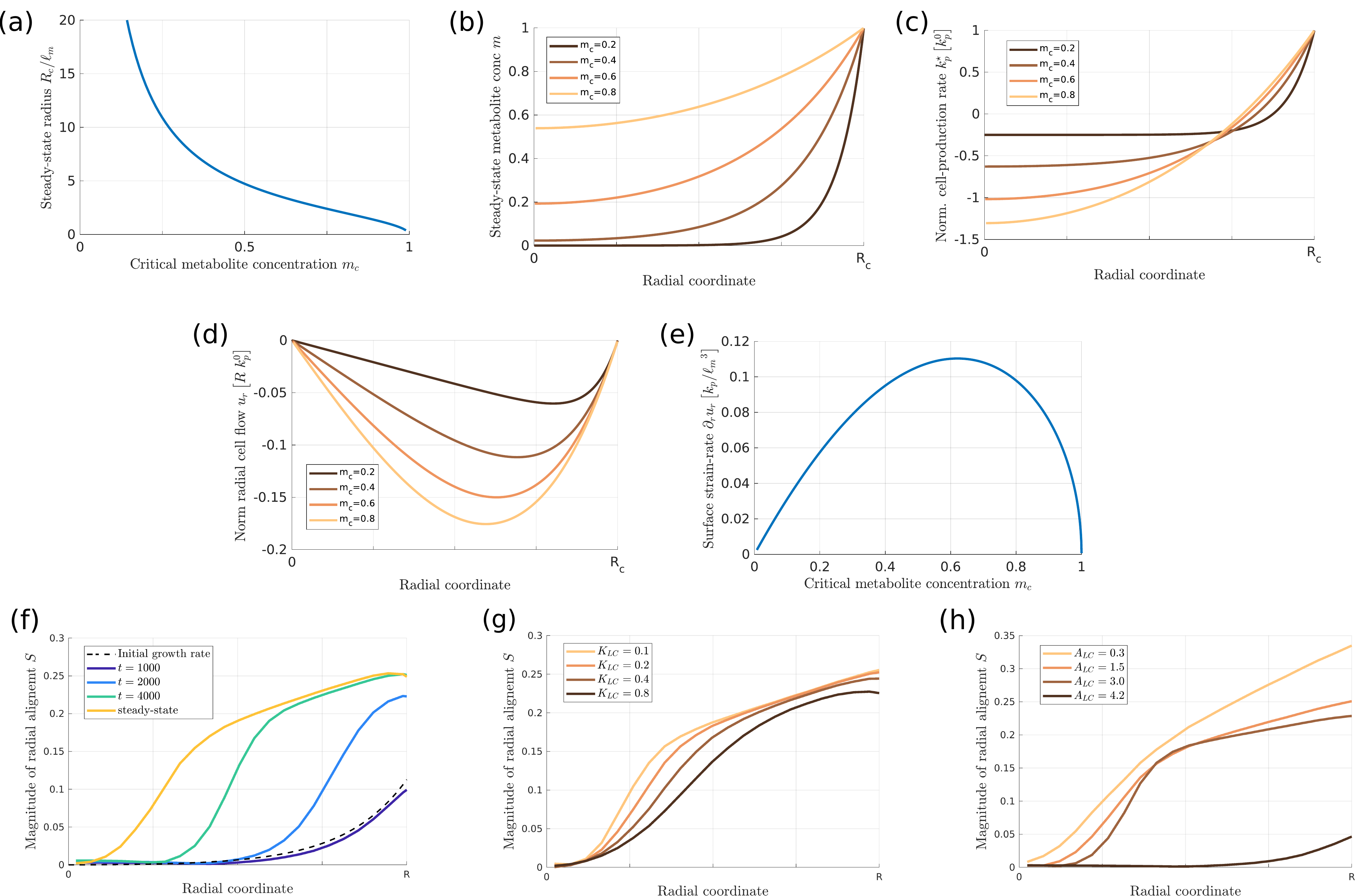}
	\caption{ \textbf{(a)} Steady-state radius $R_c/\ell_m$ as a function of critical metabolite concentration $m_c$. \textbf{(b-e)} In steady-state spheroids the critical concentration $m_c$ controls the shape of the radial metabolite profile $m$ (panel b), cell production profile $k_p^\star$ (panel c) and flow profile $u_r$ (panel d). The cell production rate $k_p^\star$ and cell flow $u_r$ are normalized to the cell production rate at the surface, $k_p^0 = k_p (1-m_c)$. \textbf{(e)} The strain rate at the surface of spheroids, which drives flow-induced cell elongation, scales as $\partial_r u_r \sim k_p \ell_m^{-3}$ and reaches a maximum value at $m_c \approx 0.6$. \textbf{(f)} Time-evolution of the radial cell alignment $Q_{rr} \sim S$ in numerical simulations initialized as $S=0$ with $\xi>0$. Initially $S$ follows the linear growth rate eqn.~(\ref{eq:Qugr}) (black, dashed line) At late times, however, non-linear contributions arising from advection and the molecular field $\mathbf{H}$ balance the growth rate and $S$ reaches a steady-state profile (yellow line). \textbf{(g,h)} Numerical cell alignment profiles at steady-state for different values of cell parameters $A_{LC}$ and $K_{LC}$. Simulations were performed with spheroids of size $R=30$ using isotropic cells ($S_{eq}=0$) and $k_p=0.001$, $\xi=0.3$.}
	\label{figS1}
\end{figure*}

 \begin{figure*}
	\centering
	\includegraphics[width=14cm]{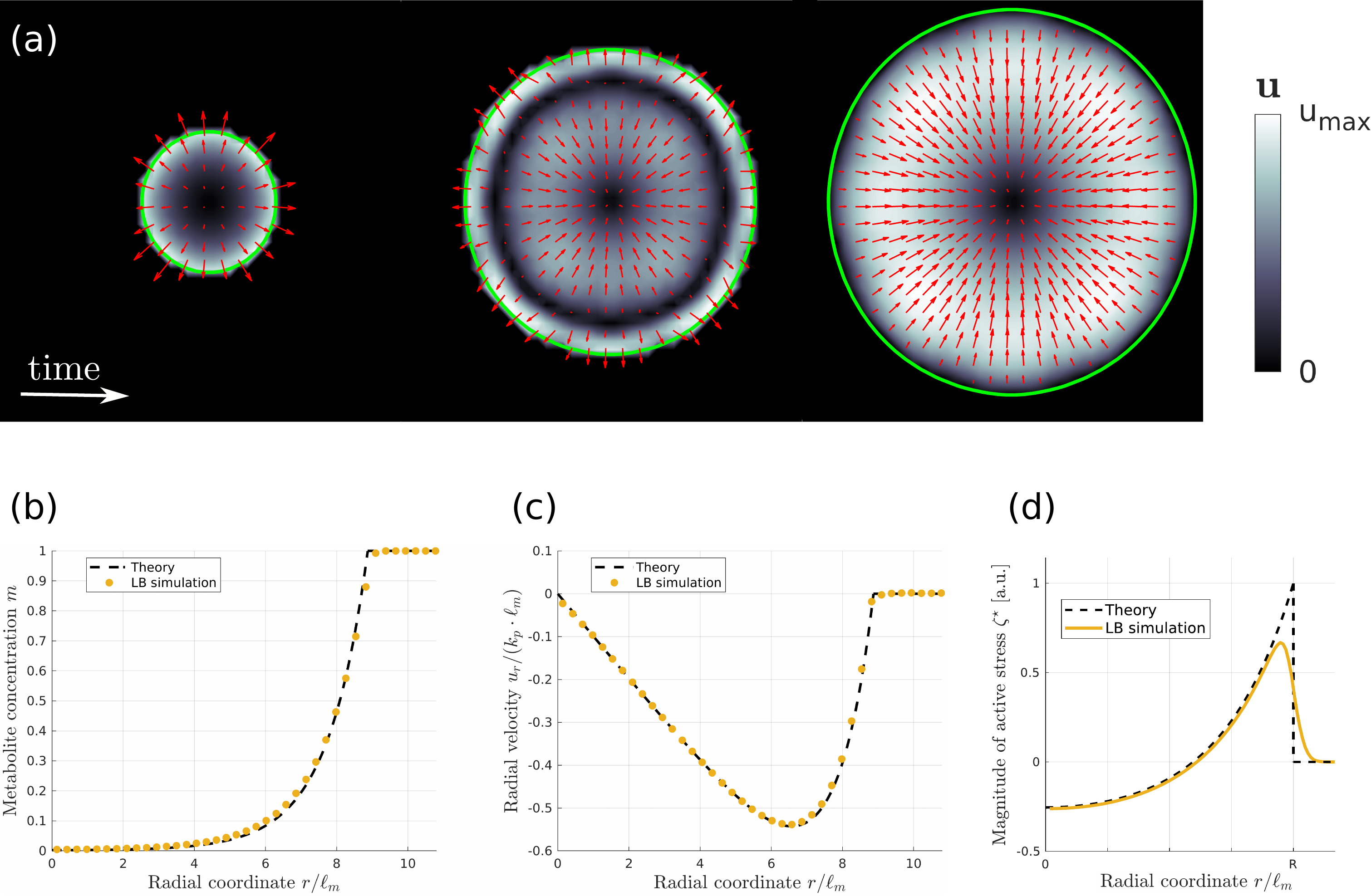}
	\caption{\textbf{(a)} Snapshots showing the time evolution of the velocity field $\mathbf{u}$ inside a growing cell aggregate ($\zeta=0$) using the hybrid lattice Boltzmann-finite difference method. At early times the spheroid is in a growing state $\dot{R}>0$ because sufficient metabolites are available throughout the aggregate, $m(r)>m_c$, creating diverging cell flows $u_r>0$. As the aggregate size increases, cells in the core will eventually have insufficient access to metabolites, $m(r=0)<m_c$, resulting in converging cell flows $u_r<0$ towards the centre. Spheroids will finally reach a steady state in which cell division and death exactly balance, $\dot{R}=0$, leading to $u_r<0$ throughout the aggregate. Snapshots show the cross-section at the equator of a growing spheroid defined by a concentration field $\varphi$ (see appendix \ref{appx:eom}), where the boundary of the aggregate is marked by the contour $\varphi = 0.5$ (green line). \textbf{(b-d)} Comparison between metabolite concentration $m$, velocity field $u_r$ and active stress $\zeta^\star$ obtained in lattice Boltzmann simulations (orange) and analytical solutions (black, dotted line) following eqns.~(\ref{eq:msphere}, \ref{eq:ur}). We validated that in the absence of sources and sinks in the continuity equation, $\mathbf{\nabla} \cdot \mathbf{u} = 0$, the flow fields obtained from the lattice-Boltzmann solver remain incompressible, with density fluctuations $\delta \rho/\rho < 0.03$.}
	\label{figS2}
\end{figure*}

 \begin{figure*}
	\centering
	\includegraphics[width=18cm]{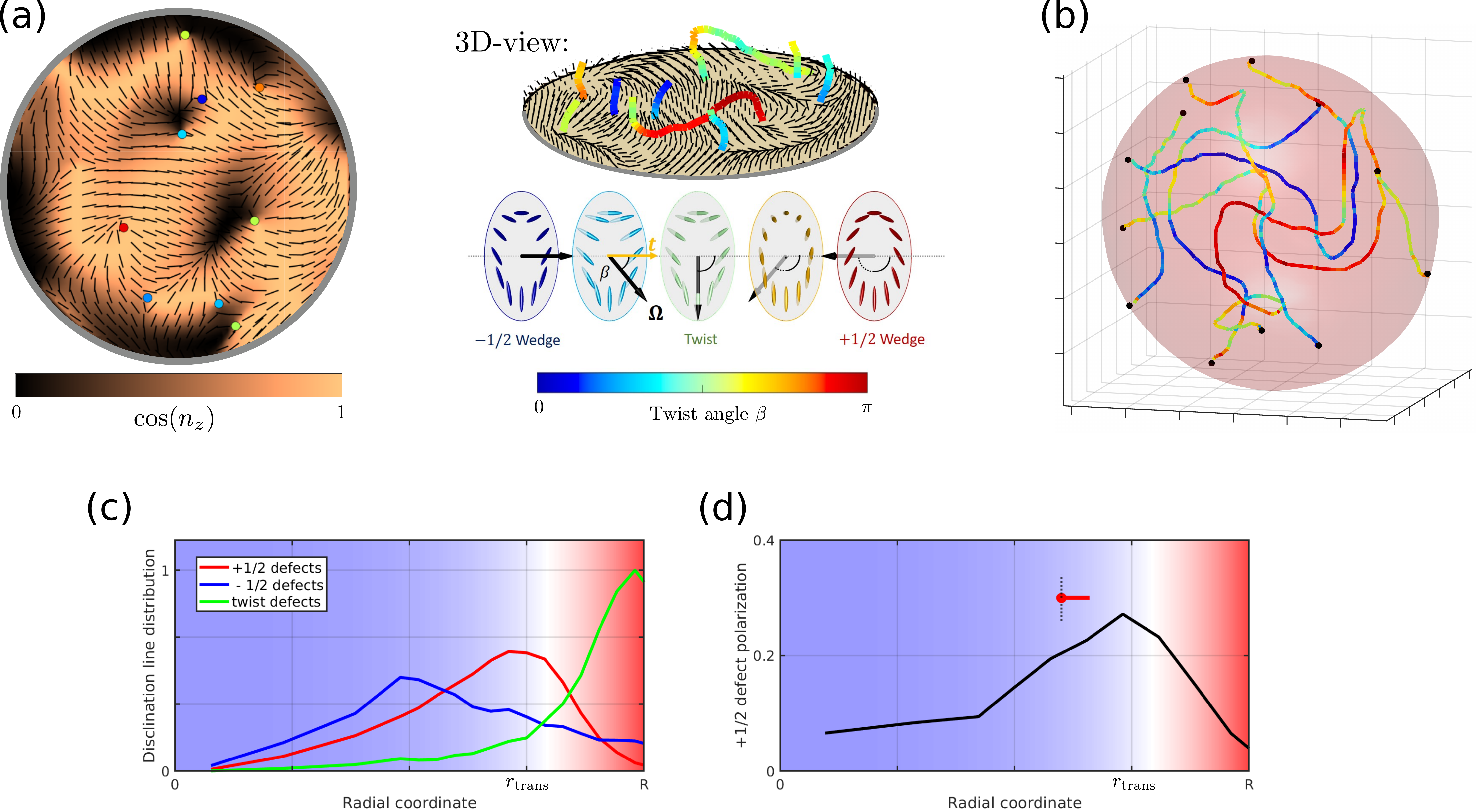}
	\caption{\textbf{(a)} In two-dimensional cross-sections of the 3D director field inside active spheroids one can identify topological defects as points where the nematic director field discontinuously changes its direction. If the director field around a defect has no out-of-plane component, $n_z=0$, it is called a wedge-type disclination and resembles $\pm 1/2$ defects in 2D systems. If the out-of-plane component $n_z$ {significantly varies} around a defect it is called a twist-type disclination. Disclination lines in three dimensions can continuously transform from a local $-1/2$ configuration (in the plane perpendicular to the line) into a $+1/2$ configuration through an intermediate twist winding. Disclination lines can be locally classified by the \textit{twist-angle} $\beta$ between the axis $\mathbf{\Omega}$ that the director field winds around and the local line tangent $\mathbf{t}$ (yellow arrow). \textbf{(b)} Due to activity disclination lines act as self-propelled entities moving through the fluid leading to spatiotemporally chaotic flows. Disclination lines constantly undergo transformation events such as breakup, recombination, nucleation and annihilation and form either closed, charge-neutral loops in the bulk or terminate at the surface of droplets. \textbf{(c)} The distribution of disclination types follows the activity profile $\zeta^\star \sim (m-m_c)$, where contractile regions in the core are dominated by wedge-type $\pm 1/2$ disclinations and twist-type defects are preferentially formed in extensile regions close to the surface. \textbf{(d)} Activity gradients in the aggregate create active torques on $+1/2$ defects which aligns them parallel to activity gradients, where the head-to-tail vector $\mathbf{p} \sim \mathbf{\nabla}\zeta^\star$. Using a polarization order parameter $\mathcal{P}$ \citep{ruske2022activity} to quantify the magnitude of defect alignment, we find that $+1/2$ defects are radially polarized throughout the aggregate. The polarization order parameter $\mathcal{P}$ reaches a maximum at the radius $r \approx r_{\text{trans}}$, where $\zeta^\star \approx 0$ and activity gradients dominate isotropic active turbulence, $\mathbf{Q} \mathbf{\nabla} \zeta^\star > \zeta^\star \mathbf{\nabla} \mathbf{Q}$. This is also reflected in the radial defect distribution in (c), where radially polarized $+1/2$ defects move outwards/inwards in contractile/extensile regions, thus accumulating around $r\approx r_{\text{trans}}$. Numerical details of the detection of disclination lines and the calculation of the twist angle $\beta$ are outlined in appendix \ref{appx:discl}.
}
	\label{figS3}
\end{figure*}

 \begin{figure*}
	\centering
	\includegraphics[width=18cm]{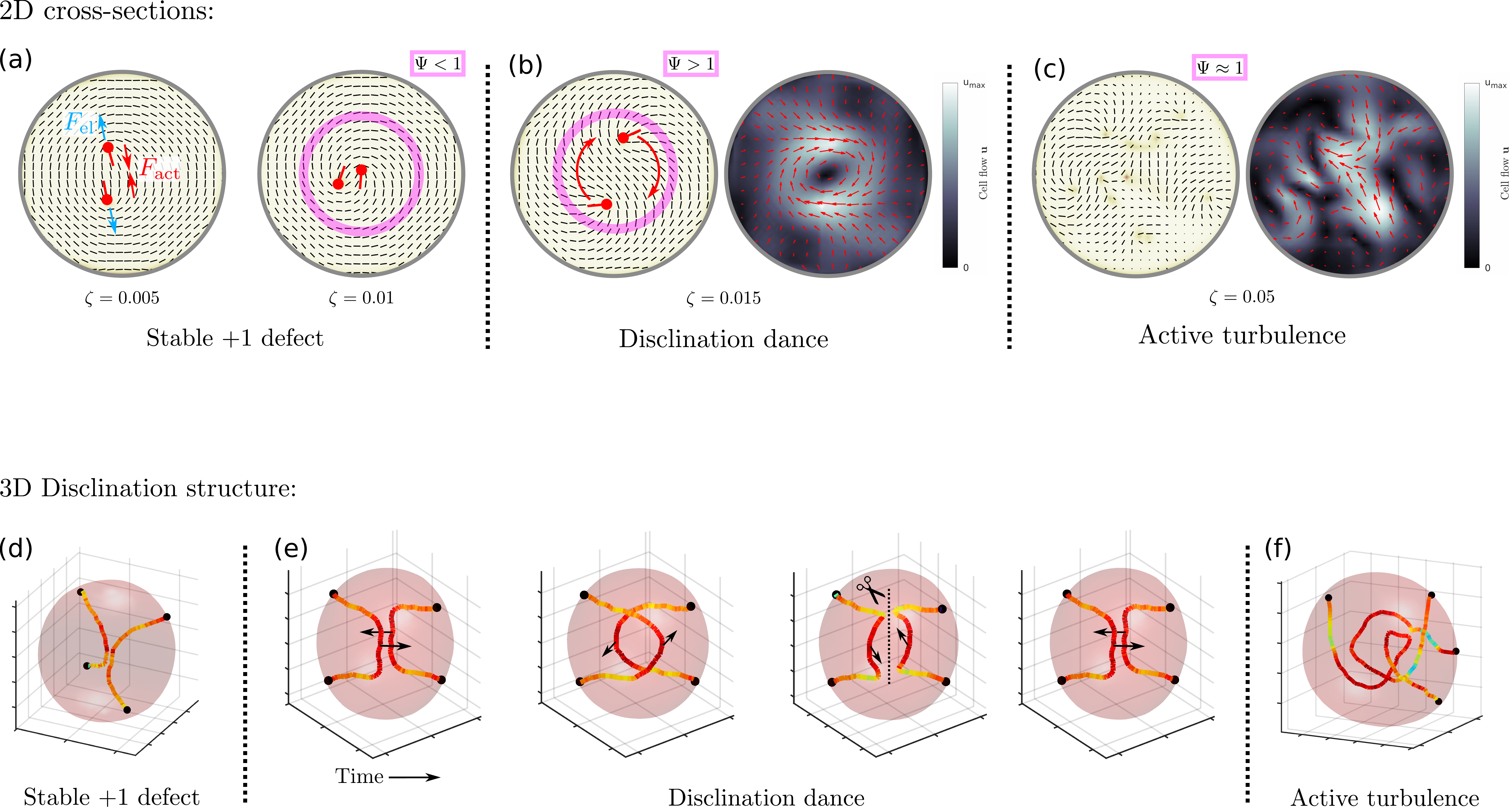}
	\caption{ As the activity $\zeta$ is progressively increased inside proliferating spheroids with $\xi<0$, the droplet undergoes three dynamical regimes: first a quiescent regime where converging flows are dominant (a,d), then a rotational \textit{dancing disclination} regime, where disclination lines spontaneously set up a rotational flow and, finally, active turbulence. \textbf{(a,d)} Two-dimensional cross-sections of the director field and disclination line structure. At very low activity the director field is stationary and the 2D cross-section resembles a $+1$ defect with angular director alignment ($\Psi<1$), where two $+1/2$ defects are located at a small finite distance to the centre. Contractile stress around defects creates active forces $F_{act}$ pointing towards the centre (red). If activity is sufficient small, active forces are balanced by repulsive elastic forces $F_{el}$ (blue) arising from an increased elastic energy as the distance between $+1/2$ defects decreases. \textbf{(b,e)} If activity surpasses a critical threshold $\zeta_c$, the system reaches a steady-state where the two $+1/2$ disclination lines start orbiting around each other, thereby creating persistent rotational motion in the core of spheroids, as shown by the cross-section of the velocity field. The outward-facing orientation of the defects creates significant radial alignment of the director field in the vicinity of the orbiting defects ($\Psi>1$). The rotational motion of defects in the core makes it inevitable that disclination lines must cross after each full rotation, thereby rewiring some line segments (see third panel in e). \textbf{(c,f)} As activity progressively increases, the motion of disclination lines and flow fields becomes more chaotic and the system eventually reaches active turbulence. The snapshots shown in this figure were obtained for $\xi=-0.4$ and disclination lines are coloured according to their local twist-angle $\beta$ using the color map shown in Fig.~\ref{figS3}.}
	\label{figS4}
\end{figure*}

\FloatBarrier

\twocolumngrid

%

\end{document}